Artificial Nonlinearity Generated from Electromagnetic Coupling Meta-molecule

Yongzheng Wen and Ji Zhou[*]

*State Key Laboratory of New Ceramics and Fine Processing, School of Materials Science and Engineering, Tsinghua University, Beijing 100084, People's Republic of China*

*Corresponding author: zhouji@tsinghua.edu.cn

**Abstract:** A purely artificial mechanism for optical nonlinearity is proposed based on a metamaterial route. The mechanism is derived from classical electromagnetic interaction in a meta-molecule consisting of a cut-wire meta-atom nested within a split-ring meta-atom. Induced by the localized magnetic field in the split-ring meta-atom, the magnetic force drives an anharmonic oscillation of free electrons in the cut-wire meta-atom, generating an intrinsically nonlinear electromagnetic response. An explicit physical process of a second-order nonlinear behavior is adequately described, which is perfectly demonstrated with a series of numerical simulations. Instead of "borrowing" from natural nonlinear materials, this novel mechanism of optical nonlinearity is artificially dominated by the meta-molecule geometry and possesses unprecedented design freedom, offering fascinating possibilities to the research and application of nonlinear optics.

Nonlinear optics has become thriving since the first observation of the second harmonic generation (SHG) in 1961 [1], and plays an essential role in many optical devices such as frequency up-converters and mixers, nonlinear spectrometers, and new light sources [2]. Over the past half century, with tons of efforts focused on searching new nonlinear materials [3, 4], researchers endeavor to uncover the physical mechanism behind the optical nonlinearity [5]. As a universal phenomenon, nonlinearity is exhibited by almost all materials interacting with sufficiently strong light [6], and various fundamental mechanisms were proposed, such as distortion of the electron cloud, relative motion of nuclei, reorientation of molecules, electrostriction effect, and thermal effect [7]. Despite the fact that these phenomenological theories extensively advanced the development of new nonlinear materials in the past several decades, they are still inadequate to present a clear physics picture and full description of the origin of the nonlinearity, so up till now it is tremendously difficult to achieve the ambitious goal of exactly predicting, rationally designing and precisely tailoring a nonlinear material.

Metamaterial, a type of artificial material allowing unprecedented control of light and exhibiting intriguing optical properties not found in nature [8-11], may offer an opportunity of realizing custom-design nonlinear properties. Various approaches were reported to introduce the nonlinearity to metamaterials, for example, engineering meta-atoms with nonlinear insertions, such as varactor diodes [12, 13]; structuring metamaterials with metal films, whose nonlinearity arises from the surface contribution [14, 15]; and combining conventional nonlinear materials as host media, such as quantum wells [16, 17]. With these methods, rapid progress on optical nonlinearity in metamaterials has been reported, including phase mismatch-free [18], electrical control [19, 20], and giant nonlinear susceptibility [21, 22]. However, the optical nonlinear responses in almost all the reported nonlinear metamaterials are actually derived from the above-mentioned external nonlinear materials and devices, and metamaterial structures play roles of enhancement on the natural nonlinearity [23], which cannot fulfill the desire of artificially designing the nonlinearity. Lapine et al made a commendable attempt to realize the structure-based artificial nonlinearity by introducing a mechanical degree of freedom to metamaterial [24], but the delay of its response time may limit its application in high frequency due to the slow mechanical displacement of the meta-atoms.

In this work, a purely artificial mechanism for optical nonlinearity based on a classical electromagnetic (EM) interaction between meta-atoms in a meta-molecule is proposed. Without the involvement of any natural nonlinear materials, an iconic second-order nonlinear behavior of the meta-molecule is described with an explicit EM coupling mechanism and perfectly verified by a series of numerical simulations. The geometric influences on the nonlinear behavior are also studied to demonstrate the ultrahigh design freedom of the artificial nonlinearity.

Taking a careful look at the classical theory of electromagnetism, we can readily reveal anharmonic motion of the electrons driven by the inherently nonlinear magnetic component of the Lorentz force, which has been neglected for a long time in either nonlinear natural crystals or metamaterials [14, 23, 25], due to the weak magnetic field in EM wave and slow drift velocity of electrons. The electric field enhancement of the split-ring resonator (SRR) has been noticed and studied in various metamaterials [26, 27], however, its enhancement on the magnetic field is comparably less-investigated [28, 29]. Based on the theoretical conception of the intrinsic nonlinearity of the magnetic force, a metamaterial comprised of EM coupling meta-molecules was designed as illustrated in Fig.1. The meta-molecule consists of two meta-atoms: a cut-wire meta-atom nested within an SRR meta-atom. With the normal incidence of an *x*-polarized EM wave, a circulating surface currents of the SRR at resonance can induce a magnetic field perpendicular to the meta-molecule plane, localized inside the SRR and dramatically enhanced by hundreds of times compared with the incident one [28]. Meanwhile, driven by the electric field of the incident wave, the free electrons in the cut-wire meta-atom move in *x* direction with a drift velocity, $\vec{v}$. As the cut wire locates inside the enhanced magnetic field, a strong magnetic force orthogonal to the drift velocity is generated. Therefore, the total force applied to the free electrons in the cut-wire meta-atom is $\vec{F}_{total} = \vec{F}_E + \vec{F}_B = q\vec{E}(\omega)e^{-i\omega t} + q\vec{v} \times \vec{B}(\omega)e^{-i\omega t} + c.c.$, where *c.c.* is complex conjugate, $q$ is elementary charge, $t$ is time, $\vec{E}(\omega)$ and $\vec{B}(\omega)$ are vectorial amplitudes of the local electric and magnetic fields at angular frequency $\omega$ respectively.

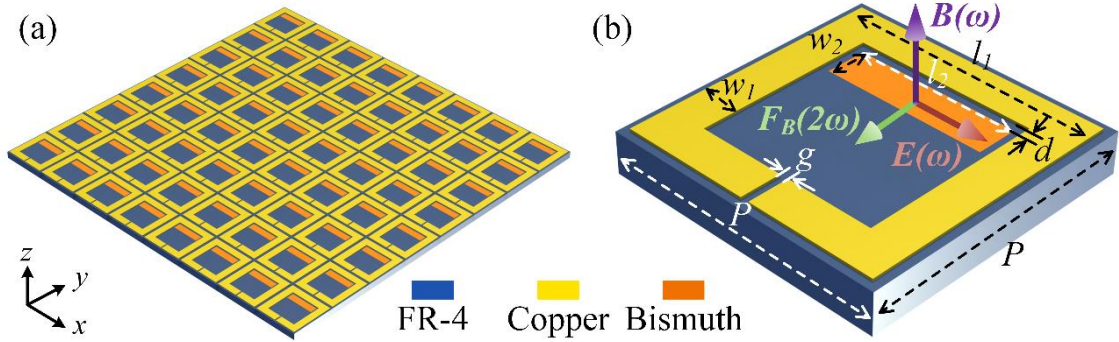

FIG. 1. Schematic of array (a) and unit cell (b) of the meta-molecules. The local magnetic and electric fields, and the magnetic force are marked. In microwave regime, the geometric constants: $l_1$=3.4 mm, $w_1$=0.45 mm, $g$=0.1 mm, $l_2$=2.3 mm, $w_2$=0.5 mm, $d$=0.1 mm, and $P$=3.6 mm.

With the EM coupling, the motion of the free electrons in the cut-wire meta-atom can be described by a modified Drude-Lorentz model as

$$m^* \frac{d^2\vec{r}}{dt^2} + m^*\gamma \frac{d\vec{r}}{dt} + m^*\omega_0^2 \vec{r} = q\vec{E}(\omega)e^{-i\omega t} + q\vec{v} \times \vec{B}(\omega)e^{-i\omega t} + c.c., \qquad (1)$$

where $\vec{r}$ is the displacement from the equilibrium position, $m^*$ is the effective electron mass, $\gamma$ is the electron collision rate, and $\omega_0$ is the angular eigenfrequency, which is 0 for metal and heavily-doped semiconductor used in our case. To demonstrate the artificial nonlinearity from the meta-molecule without losing generalities, a widely-known and first-observed nonlinear process, SHG, was studied as an example. In that case, the drift velocity of the electron is expressed as $\vec{v} = \tilde{\mu}_e \vec{E}(\omega) e^{-i\omega t}$, where $\tilde{\mu}_e$ is the mobility of the free electrons in Drude model, and the magnetic force becomes

$$\vec{F}_B = q\tilde{\mu}_e \vec{E}(\omega) e^{-i\omega t} \times \vec{B}(\omega) e^{-i\omega t} + c.c. = q\tilde{\mu}_e \vec{E}(\omega) \times \vec{B}(\omega) e^{-i2\omega t} + c.c.. \quad (2)$$

The exhibited second-order term in the magnetic force oscillates the free electrons in an anharmonic way, supplying a clear and designable mechanism of the artificial nonlinearity. However, in most materials, the magnetic force hardly plays a role due to the weak magnetic component in the EM wave, and it points in the direction of the incident wave vector, which is difficult to generate the second-harmonic wave with forward propagation. In this meta-molecule, the dramatic enhancement of the localized magnetic field in the SRR makes the magnetic force much more significant. Since the induced magnetic field along $z$ axis and the electric filed along the $x$ axis, the magnetic force oscillates the electrons in the cut wire in $y$ direction, as shown in Fig. 1(b). Hence, it can be expected that the second-harmonic wave radiates along both $+z$ and $-z$ directions in $y$ polarization. More importantly, this nonlinearity intrinsically originates from the magnetic force rather than the properties of the composites, which fundamentally distinguishes our structure with most reported nonlinear metamaterials [20, 23, 30]

To physically describe the SHG from the meta-molecule and derive the nonlinear polarization, the perturbation method is used to solve the Eq. (1), which is evolved into a second-order differential equation by substituting the Eq. (2). A displacement with first and second harmonic terms is assumed as a general solution, $\vec{r} = \vec{r}_1 + \vec{r}_2 = \vec{r}_1(\omega) e^{-i\omega t} + \vec{r}_2(2\omega) e^{-i2\omega t} + c.c.$. We separate the first and second order terms and rearrange the Eq. (1) as

$$m^* \frac{d^2 \vec{r}_1(\omega) e^{-i\omega t}}{dt^2} + m^* \gamma \frac{d\vec{r}_1(\omega) e^{-i\omega t}}{dt} = q\vec{E}(\omega) e^{-i\omega t}, \quad (3a)$$

$$m^* \frac{d^2 \vec{r}_2(2\omega) e^{-i2\omega t}}{dt^2} + m^* \gamma \frac{d\vec{r}_2(2\omega) e^{-i2\omega t}}{dt} = q\tilde{\mu}_e \vec{E}(\omega) \times \vec{B}(\omega) e^{-i2\omega t}. \quad (3b)$$

For clarity, the complex conjugate is eliminated. Eq. (3a) presents similar linear form with the classical Drude model, and its solution is

$$\vec{r}_1(\omega) = -\frac{q}{m^*} G(\omega) \vec{E}(\omega), \quad (4)$$

where

$$G(\omega) = \frac{1}{\omega^2 + i\omega\gamma}. \quad (5)$$

By solving the nonlinear equation Eq.(3b), we can obtain:

$$\vec{r}_2(2\omega) = -\frac{q\tilde{\mu}_e}{m^*} G(2\omega) |\vec{B}(\omega)||\vec{E}(\omega)| \hat{a}_y, \quad (6)$$

where $\hat{a}_y$ is the unit vector along the $y$ axis to indicate orientation of the nonlinear displacement driven by the magnetic force. With the Eq.(4) and Eq.(6), the first and second order polarizations ($\vec{P}_L(\omega)$ and $\vec{P}_{NL}(2\omega)$) can be calculated as

$$\vec{P}_L(\omega) = -\varepsilon_0 \omega_p^2 G(\omega) \vec{E}(\omega), \quad (7a)$$

$$\vec{P}_{NL}(2\omega) = -\frac{\varepsilon_0 \omega_p^2 \mu_{e0}}{1 - i\omega/\gamma} |\vec{B}(\omega)||\vec{E}(\omega)| G(2\omega) \hat{a}_y, \quad (7b)$$

where $\varepsilon_0$ is the vacuum permittivity, and $\omega_p$ and $\mu_{e0}$ are the plasma frequency and the dc mobility of the material forming the cut-wire meta-atom, respectively. The mathematical details are described in section I in the Supplemental Material[31].

As mentioned above, $\vec{B}(\omega)$ is originally induced by the incident electric field of the EM wave. Hence, according to the Ampere's circuital law, $\vec{B}(\omega) \propto \vec{J}(\omega) = \sigma \vec{E}_0(\omega)$, where $\vec{E}_0(\omega)$ is the incident electric field, $\sigma$ is the conductivity of the SRR and $\vec{J}(\omega)$ is the surface current density in the SRR. Considering the fact that the strength of the local electric field is basically the same as the incident one, it can be conveniently derived from the Eq. (7b) that the nonlinear polarization is in a quadratic relation with the incident electric field, $|\vec{P}_{NL}(2\omega)| \propto |\vec{E}_0(\omega)|^2$, which is a signature phenomenon of the second-order nonlinearity. Meanwhile, the high conductivity of the material composing the SRR would benefit the SHG intensity as well, since it increases the current density and provides stronger magnetic field.

As seen from the Eq. (7b), improving the mobility of the cut-wire meta-atom would also strengthen the SHG proportionally. It is because driven by the same electric field, the free electrons would gain higher drift velocity with the better mobility, leading to a more significant magnetic force.

To verify our theoretical prediction on the artificial second-order nonlinearity, the meta-molecule with the same structure shown in Fig.1 was modeled and simulated by a commercial finite-element package (COMSOL Multiphysics), and its geometrical constants were optimized to work in the microwave regime and provide high SHG intensity. Single unit cell was simulated with the periodic boundary (see section II in the Supplemental Materials[31]). The substrate was 1 mm thick FR-4 with the permittivity of 4.2+0.1$i$. Guided by the theory, high conductivity of the SRR and high mobility of the cut wire would both benefit the SHG. Accordingly, the SRR is comprised of a 30 μm thick copper layer with the conductivity of 4.5×10$^7$ S/m [35], and a 100 nm thick bismuth film is modeled as the cut wire with the conductivity ($\sigma_0$) of 2.2×10$^5$ S/m and the mobility ($\mu_{Bi}$) of 0.11 m$^2$/V•s [36, 37]. All compositions are treated as linear materials in the simulation. To take the localized magnetic field into account, the bismuth film was modelled with an anisotropic conductivity tensor ($\sigma(\omega)$) derived from a rigorous definition of conductivity including the term of magnetic field (see section III in the Supplemental Material[31]), as follows [38, 39]

$$\sigma(\omega) \approx \sigma_0 \begin{bmatrix} \frac{1}{1+(\mu_{Bi}B)^2} & -\frac{\mu_{Bi}B}{1+(\mu_{Bi}B)^2} & 0 \\ \frac{\mu_{Bi}B}{1+(\mu_{Bi}B)^2} & \frac{1}{1+(\mu_{Bi}B)^2} & 0 \\ 0 & 0 & 1 \end{bmatrix}, \quad (8)$$

where $B$ is the local magnetic field inside the SRR. Despite the utilization of bismuth in this specific implementation, the anisotropic model of conductivity is general and applies to all conductors with the presence of the magnetic field. It should be noted that the bismuth is chosen solely due to its high conductivity and mobility, which enhance the artificial nonlinear response of the meta-molecule. As the nonlinearity arises from the meta-molecule structure rather than the compositions, other conductors, such as doped silicon, could also be used (see section IV in the Supplemental Material[31]).

Under the normal illumination of an $x$-polarized plane wave from the top, the reflection, transmission and absorption spectra of the meta-molecule were first simulated and plotted in Fig. 2(a). The localized magnetic field reaches its maximum at 10 GHz, which is slightly lower than the resonant frequency of 10.8 GHz. Fig. 2(b) shows the surface currents and magnetic field distributions at 10 GHz of the unit cell, and the circulating currents produce the enhanced magnetic field as maximum 79.5 times stronger as the incident one. The existence of the cut wire has no observable influence on the resonance of the SRR, and the induced magnetic field can penetrate through the cut wire due to its thinness.

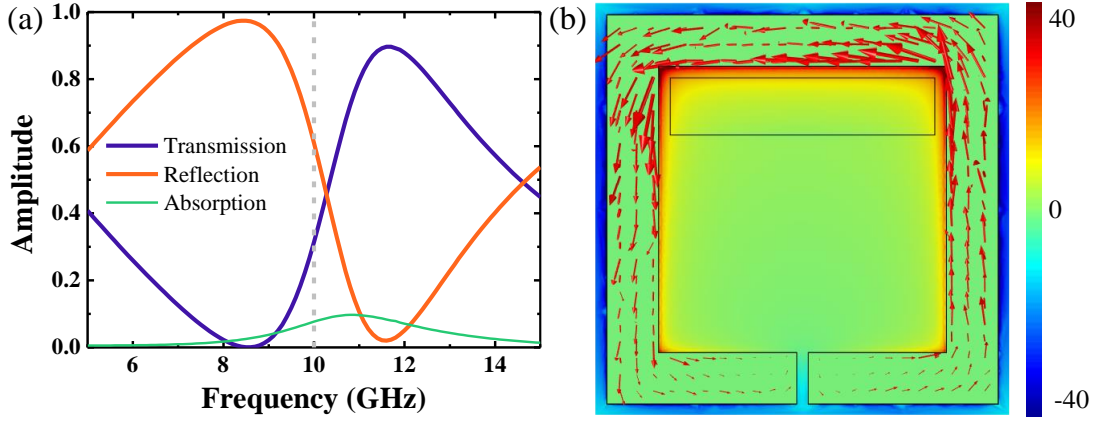

FIG. 2. Transmission, reflection and absorption spectra of the meta-molecule (a) with 10 GHz marked with grey dashed line; the magnetic field distribution (b) with the scale normalized to the incident magnetic field amplitude and the orientation of the surface currents marked with red arrows.

The time-domain response of the meta-molecule was then simulated with a Gaussian pulsed plane wave at 10 GHz casted from the top, and the nonuniformity of the localized magnetic field was taken into account. The peak intensity of the incident electric field was $1\times10^7$ V/m, and the strength of the magnetic force on the free electrons of the cut-wire meta-atom can be theoretically calculated as 29.15% of that of the electric force, which is significant enough to generate a nonlinear response (see section V in the Supplemental Material[31]). As revealed in Fig. 3(a), we examined a $y$-polarized transmission spectrum in time domain, and a 20 GHz wave can be extracted by high-pass filtering with the peak electric field of 613.7 V/m. Compared with the incident fundamental wave, the doubled frequency demonstrates the proposed meta-molecule successfully generates the second-harmonic wave. Due to the weak electric field in the tail bounds of the Gaussian pulse, the pulse width of the SHG signal is narrower than that of the incident wave, resembling the natural nonlinear materials. The frequency spectrum in Fig. 3(b), transformed from the time spectrum by Fourier transformation, exhibits the SHG more obviously. We also observed the $x$-polarized transmission time and frequency spectra (not shown) but found no SHG. All these simulated results agree perfectly with the theory.

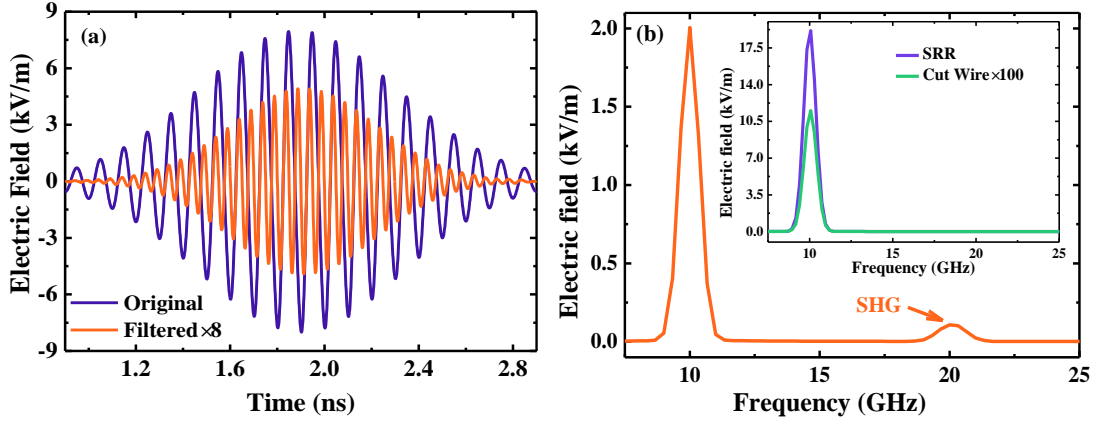

FIG. 3. Original and high-pass filtered time-domain transmission spectra of the nonlinear meta-molecule in *y* polarization (a) and its frequency-domain spectrum (b), the inset is the frequency spectra of two meta-atoms.

To elucidate the necessity of the EM coupling between the two meta-atoms for the artificial nonlinearity, the metamaterials containing only SRR or cut-wire meta-atoms were simulated with the same incident Gaussian pulsed plane wave at 10 GHz, and the copper of the SRR was modeled with the same anisotropic conductivity tensor and the mobility of $2.25 \times 10^{-3}$ m$^2$/V·s [40]. Their simulated frequency spectra are revealed in the inset of the Fig. 3(b), and there is only a fundamental wave observed and no evident SHG detected. The simulated result of SRR also supports the previously-reported hydrodynamic theory that the magnetic contribution to the nonlinear response of the metal SRR can be neglected [25].

As specified in the theory, the proposed meta-molecule radiates longitudinally in both directions. Thereby, we examined its *y*-polarized reflection spectra in frequency and time domains. As shown in Fig.4, the SHG is also distinct with the peak electric field of 453.1 V/m. Together with the transmitted one, the total conversion efficiency of the meta-molecule is calculated as $5.8 \times 10^{-9}$, corresponding to an effective second-order nonlinear susceptibility of $1.2 \times 10^3$ pm/V (see section VI in the Supplemental Material[31]). This nonlinear susceptibility is not only highly competitive to the reported nonlinear metamaterials based on metal plasmonic resonators [41], but comparable with the traditional nonlinear crystal [6, 7].

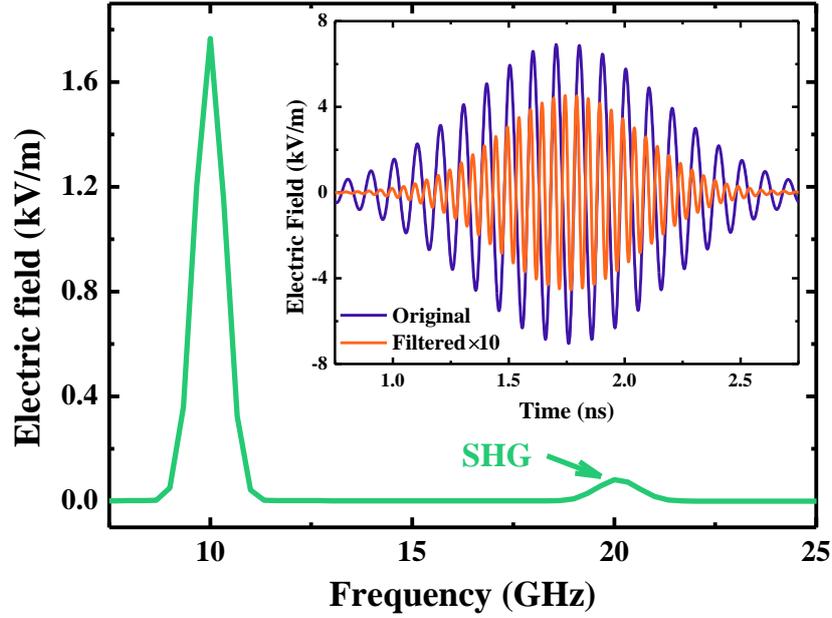

FIG. 4. Reflection frequency spectrum of the meta-molecule in *y* polarization, and the inset is its original and high-pass filtered time-domain spectra.

To confirm the nonlinearity generated from the meta-molecule is capable of artificial manipulation, we studied the relations between the intensity of the transmitted SHG and the two geometric constants of the structure: the distance between two meta-atoms (*d*) and the width of the cut-wire meta-atom ($w_2$). Since the nonlinear response is proportional to the magnetic field, which is nonuniformly distributed inside the SRR as depicted in Fig. 2(b), the narrower distance between two meta-atoms locates the cut wire in the region with stronger magnetic field and leads to higher SHG intensity, especially in the range less than 0.1 mm, as plotted in Fig.5. The inset of Fig.5 shows the other geometric impact and presents the increase of the cut wire width enhances the strength of the SHG in a linear manner. It is easy to understand that the wider resonator means more free electrons can interact with the localized magnetic field, thus leading to a stronger nonlinear response. Inspired by these two examples, it can be conveniently predicted that other geometric constants, which would influence the resonant behavior of the meta-molecule, could also have an effective impact on the nonlinear behavior, including the gap (*g*) and length ($l_1$) of the SRR, and the lattice constant (*P*).

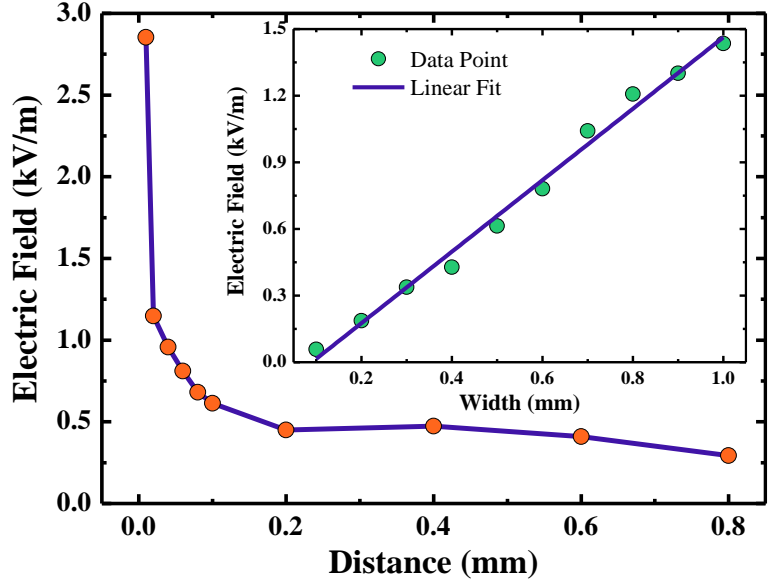

FIG. 5. Relation between the intensity of the transmitted SHG and the distance between two meta-atoms. Inset: the intensity of the SHG with different width of the cut-wire meta-atom, and the linear fit of the data points.

All these simulated results adequately verify the proposed mechanism of artificial nonlinearity without involving any sophisticated mesoscopic and quantum mechanisms in materials or low frequency mechanical process. The SRR with cut-wire structure may also play a role in some previously-reported nonlinear metamaterials. For example, by designing a split-ring slit combined with a cut-wire slit, Ren et al pave an innovative road of realizing unnaturally giant nonlinear optical activity, in which the origin of the optical nonlinearity still resides in gold film instead of the structure [22]. In the design proposed here, the optical nonlinearity essentially arises from a magnetic-force-based EM coupling process, dominated by the meta-molecule structure, rather than the electric-field-enhanced natural nonlinearity in most reported work. One convincing proof of the uniqueness of this mechanism is that as described in the physical model above, the second-harmonic wave is actually generated from the cut-wire meta-atom, which is in sharp contrast to the consensus generally held by others that the centrosymmetric cut wire cannot provide SHG [25, 42, 43].

Instead of "borrowing" from nature, this artificial mechanism offers metamaterials the ability of generating the optical nonlinearity themselves. More importantly, by simply structuring the inclusion geometry of the metamaterials, the artificial optical nonlinearity can be precisely designed with unprecedented freedom, which would bring unlimited possibilities and myriad novel phenomena to the nonlinear optics. Some quick examples may include the flat nonlinear lens and nonlinear holography. Although this proof-of-concept meta-molecule is numerically demonstrated in the microwave regime, due to the wide applicability of classical electromagnetism, this theory of artificial nonlinearity can be easily extended to higher frequencies, such as terahertz and infrared (see section VII in the Supplemental Material[31]). Meanwhile, the meta-molecule is highly feasible in experiment. The microwave and terahertz meta-molecules could be conveniently manufactured by microfabrication techniques with standard ultraviolet lithography, and the infrared sample could be processed by nanotechnology with electron-beam lithography. All the metal compositions, such as copper, gold and bismuth, could be deposited by electron-beam evaporation and magnetron sputtering, and some conductive materials with high mobility may also be involved as the cut-wire meta-atom, such as graphene and InAs.

In conclusion, we theoretically demonstrated a novel mechanism for purely artificial optical nonlinearity generated from a meta-molecule consisting of two EM coupled meta-atoms. Interacted with the magnetic field localized in the SRR meta-atom, the free electrons in the cut-wire meta-atom oscillate in anharmonic way under the magnetic force, which generates the nonlinear response. Based on the classical electromagnetism, the physical process of an iconic second-order nonlinearity of the meta-molecule is explicitly described and perfectly supported by the numerical simulations. The geometric influences on the nonlinear behavior demonstrate this innovative mechanism possesses ultrahigh degree of design freedom. This purely EM mechanism of artificial nonlinearity, without involvement of any photo-induced electronic, thermal, mechanical, or quantum processes in natural materials, supplies a metamaterial-based approach for designing the nonlinear optical materials, which would open a wide range of possibilities and bring fantastic potentials to nonlinear optics.

This work was supported by the National Natural Science Foundation of China under Grant Nos. 11274198 and 51532004 and the China Postdoctoral Science Foundation under Grant No. 2015M580096.

Supplemental Material:

Artificial Nonlinearity Generated from Electromagnetic Coupling Meta-molecule

Yongzheng Wen and Ji Zhou[*]

*State Key Laboratory of New Ceramics and Fine Processing, School of Materials Science and Engineering, Tsinghua University, Beijing 100084, People's Republic of China*

[*]Corresponding author: zhouji@tsinghua.edu.cn

I. MATHEMATICAL DETAILS OF POLARIZATION EQUATIONS.

For the convenience of the mathematical derivation, the Eq.(4) and (6) in the main text are repeated here

$$\vec{r}_1(\omega) = -\frac{q}{m^*} G(\omega) \vec{E}(\omega), \qquad (S1a)$$

$$\vec{r}_2(2\omega) = -\frac{q\tilde{\mu}_e}{m^*} G(2\omega) |\vec{B}(\omega)| |\vec{E}(\omega)| \hat{a}_y, \qquad (S1b)$$

where

$$G(\omega) = \frac{1}{\omega^2 + i\omega\gamma}. \qquad (S2)$$

Every symbol in the equations has been defined in the main text. As the polarization vector ($\vec{P}$) is the density of the dipole moments, it can be expressed as $\vec{P} = Nq\vec{r}$, where $N$ is the free electron density. Thus, the first-order polarization ($\vec{P}_L(\omega)$) can be obtained as

$$\vec{P}_L(\omega) = Nq\vec{r}_1(\omega) = -\frac{Nq^2}{m^*} G(\omega) \vec{E}(\omega) = -\varepsilon_0 \omega_p^2 G(\omega) \vec{E}(\omega). \qquad (S3)$$

where the plasma frequency is

$$\omega_p = \sqrt{\frac{Nq^2}{m^*\varepsilon_0}}. \qquad (S4)$$

Similarly, the second-order polarization ($\vec{P}_{NL}(2\omega)$) can be obtained

$$\vec{P}_{NL}(2\omega) = Nq\vec{r}_2(2\omega) = -\frac{Nq^2\tilde{\mu}_e}{m^*}G(2\omega)|\vec{B}(\omega)||\vec{E}(\omega)|\hat{a}_y$$

$$= -\varepsilon_0\omega_p^2\tilde{\mu}_e G(2\omega)|\vec{B}(\omega)||\vec{E}(\omega)|\hat{a}_y. \qquad (S5)$$

Meanwhile, since the magnetic force is always orthogonal to the direction of the drift motion and has no impact on the kinetic energy of the free electron, the modulus of the drift velocity is determined by the electric force, leading to

$$\vec{v} = \frac{d\vec{r}_1(\omega)e^{-i\omega t}}{dt} = -i\omega\vec{r}_1(\omega)e^{-i\omega t}. \qquad (S6)$$

Therefore, the mobility in Drude model can be calculated as

$$\tilde{\mu}_e = \frac{\vec{v}}{\vec{E}(\omega)e^{-i\omega t}} = \frac{\mu_{e0}}{1-i\omega/\gamma}, \qquad (S7)$$

where the dc mobility is $\mu_{e0}=q/(m^*\gamma)$. By substituting the Eq. (S7), the Eq. (S5) can be expressed as

$$\vec{P}_{NL}(2\omega) = -\frac{\varepsilon_0\omega_p^2\mu_{e0}}{1-i\omega/\gamma}|\vec{B}(\omega)||\vec{E}(\omega)|G(2\omega)\hat{a}_y. \qquad (S8)$$

The Eqs. (S3) and (S8) are the Eqs. (7a) and (7b) in the main text respectively.

II.     SIMULATION SETTINGS

As described in the main text, the meta-molecule was simulated in both frequency and time domains. In both domains, single unit cell was simulated with the periodic boundary condition in *x* and *y* directions, and the ports for transmitting and receiving wave were set on the top and bottom boundaries respectively. The localized magnetic field was incorporated by simply inputting the variable representing the local magnetic field when defining the anisotropic materials.

In frequency domain, the S-parameters of the meta-molecule were simulated ranging from 5 GHz to 15 GHz with the step of 0.1 GHz. Therefore, the reflection (*R*), transmission (*T*) and absorption (*A*) spectra can be obtained with the relations $R=|S_{11}|^2$, $T=|S_{21}|^2$ and *A=1-T-R*.

In time domain, the *x*-polarized incident wave was defined by the electric filed $\vec{E}=(\vec{E}_x, 0, 0)$, and the expression of $\vec{E}_x$ is

$$\vec{E}_x = \vec{E}(\omega)\cos(\omega t - k_0 z)e^{-(\frac{t-t_0}{\Delta t})^2}, \quad (S9)$$

where $\vec{E}(\omega)$ is the peak amplitude of the electric field, $\omega$ is the angular frequency, $t$ is time, $k_0$ is the wavenumber at the frequency $\omega$, $z$ is the coordinate in $z$ axis, and $t_0$ and $\Delta t$ are the parameters describing the Gaussian pulse. The values used in the time-domain simulations are $|\vec{E}(\omega)|=10^7$ V/m, $\omega=2\pi\times10^{10}$ rad/s, $t_0=1.6$ ns, $\Delta t =600$ ps. The total time of 4 ns was simulated with the step of 1 ps.

### III. DEVIATION OF ANISOTROPIC CONDUCTIVITY TENSOR

A rigorous definition of conductivity should include the presence of the magnetic field, because the magnetic force would also influence the motion of the electrons in the conductors. In most cases, since the ambient geomagnetic field or the magnetic field of the electromagnetic wave is too weak to cause a noticeable effect, the magnetic term in the conductivity is negligible. However, in the cases with the presence of the strong magnetic field, such as the strongly enhanced localized magnetic field in the proposed meta-molecule, the impact of the magnetic force on the motion of the electrons is too distinct to be neglected, and a second-order tensor should be used to describe the conductivity.

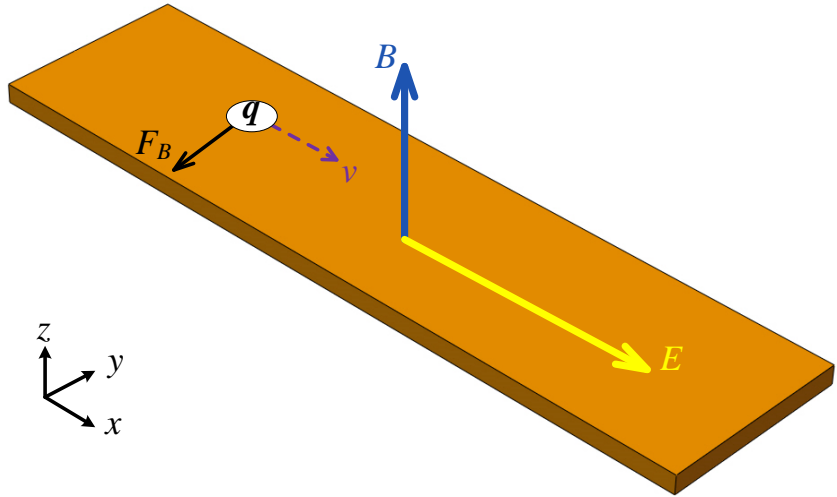

FIG. S1 A simplified model for the deviation of anisotropic conductivity tensor.

A simplified model is considered in the following deviations as shown in Fig. S1. A conductive cut-wire structure is applied with a dynamic magnetic field along $z$ axis and dynamic electric field along $x$ axis at frequency $\omega$. In this case, according to the

Drude model, the current density of the conductor, $\vec{J}$, is

$$\frac{d\vec{J}}{dt} + \gamma \vec{J} = \frac{Nq^2 \vec{E}}{m^*} + \frac{q}{m^*} \vec{J} \times \vec{B}. \qquad (S10)$$

Assuming the electric field as $\vec{E} = \vec{E}(\omega)e^{-i\omega t}$ and the current density as $\vec{J} = \vec{J}(\omega)e^{-i\omega t}$, the Eq. (S10) can be solved as

$$\vec{J}(\omega) = \frac{\sigma_0 \vec{E}(\omega)}{1 - i\omega/\gamma} + \frac{\mu_{e0}}{1 - i\omega/\gamma} \vec{J}(\omega) \times \vec{B}, \qquad (S11)$$

where $\sigma_0 = Nq^2/(m^*\gamma)$ and $\mu_{e0} = q/(m^*\gamma)$ are the isotropic dc conductivity and mobility, respectively. The isotropic mobility can also be rewritten in a tensor form as

$$\boldsymbol{\mu_{e0}} = \mu_{e0} \begin{bmatrix} 1 & 0 & 0 \\ 0 & 1 & 0 \\ 0 & 0 & 1 \end{bmatrix}. \qquad (S12)$$

The tensor form of magnetic field $\vec{B}$ can be described as [1]

$$\vec{\mathbf{B}} = \begin{bmatrix} 0 & -B & 0 \\ B & 0 & 0 \\ 0 & 0 & 0 \end{bmatrix}. \qquad (S13)$$

To almost all the conductive materials, $\omega \ll \gamma$ at the target frequency of 10 GHz, we assume $1 - i\omega/\gamma \approx 1$. By substituting the two tensors into the Eq. (S11) and rearranging it into the form of $\vec{J}(\omega) = \sigma(\omega)\vec{E}(\omega)$, we can obtain the anisotropic conductivity tensor, $\sigma(\omega)$, as

$$\sigma(\omega) \approx \sigma_0 \begin{bmatrix} \frac{1}{1 + (\mu_{e0}B)^2} & -\frac{\mu_{e0}B}{1 + (\mu_{e0}B)^2} & 0 \\ \frac{\mu_{e0}B}{1 + (\mu_{e0}B)^2} & \frac{1}{1 + (\mu_{e0}B)^2} & 0 \\ 0 & 0 & 1 \end{bmatrix}. \qquad (S14)$$

As the deviation processes start from a widely applicable Drude model and do not involve any assumption that restricts the application scope to some specified materials, the anisotropic model of the conductivity is general and applies to most conductors with the presence of the magnetic field.

IV. SIMULATION WITH DOPED SILICON AS CUT-WIRE META-ATOM

To clarify the bismuth is not necessary for the artificial nonlinearity, the same meta-molecule was simulated in microwave regime. The material for cut-wire meta-atom was replaced with n-doped silicon, which presents higher carrier mobility than the p-doped one. To make the conductivity of the silicon similar with that of the bismuth, the silicon was set as heavily-doped with the doping concentration of $2\times10^{26}$ m$^{-3}$, corresponding to the dc conductivity of $2.4\times10^{5}$ S/m and the mobility of $7.4\times10^{-3}$ m$^2$/V•s [2]. The doped silicon was modelled with the anisotropic conductivity tensor as well. All the other compositions and geometric constants were kept the same with structure shown in Fig. 1 in the main text, and the simulation settings were also kept same as described in Section II of the Supplemental Material.

The *y*-polarized transmission spectrum in time domain is revealed in Fig. S2(a), and a 20 GHz wave can be extracted by high-pass filtering with the peak electric field of 46.7 V/m. The frequency spectrum in Fig. S2(b), transformed from the time spectrum by Fourier transformation, exhibits the SHG more obviously. The simulated results clarify that the nonlinear response of the meta-molecule arises from the artificial structure rather than the compositions, and the bismuth is chosen solely due to its high conductivity and mobility, which enhance the nonlinear response. With the same structure and similar conductivity, since the bismuth offers much higher mobility than the heavily-doped silicon, the sample with bismuth presents stronger nonlinear response than that with the doped silicon, which has been clearly indicated in the theoretical model in the main text.

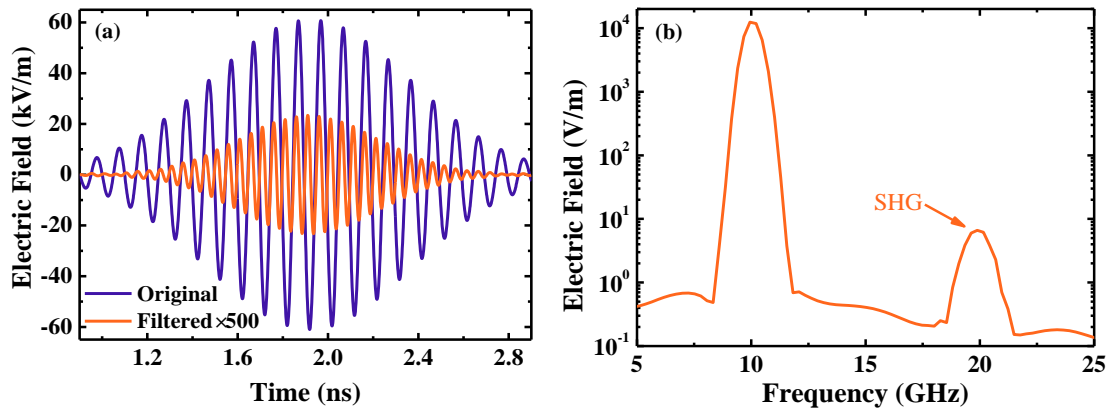

FIG. S2 Original and high-pass filtered time-domain transmission spectra in *y* polarization of the meta-molecule with doped silicon as cut-wire meta-atom (a) and its frequency-domain spectrum in logarithmic coordinate (b).

## V.   RATIO OF MAGNETIC FORCE TO ELECTRIC FORCE

As mentioned in the main text, the free electrons in the cut-wire meta-atom are under magnetic force and electric force simultaneously. We define $K$ as the ratio of the amplitudes of the two forces and obtain

$$K = \frac{|\vec{F}_B|}{|\vec{F}_E|} = \frac{q|\vec{v}||\vec{B}(\omega)|}{q|\vec{E}(\omega)|} = \frac{|\vec{v}||\vec{B}(\omega)|}{|\vec{E}(\omega)|}. \qquad (S15)$$

As the local magnetic field $\vec{B}(\omega)$ is enhanced by $N$ times compared to the incident $\vec{B}_0(\omega)$, and the drift velocity has $|\vec{v}| = \mu_{e0}|\vec{E}(\omega)|$ in microwave regime, the Eq. (S15) evolves into

$$K = \frac{\mu_{e0}|\vec{E}(\omega)|N|\vec{B}_0(\omega)|}{|\vec{E}(\omega)|} = \frac{\mu_{e0}N|\vec{E}_0(\omega)|}{c}, \qquad (S16)$$

where the relation $|\vec{B}_0(\omega)| = |\vec{E}_0(\omega)|/c$ is applied and $c$ is the speed of light. By substituting the values from the main text into the Eq. (S16), including $N$=79.5, $\mu_{e0}$=0.11 m$^2$/V•s, $E_0$=10$^7$ V/m, $c$=3×10$^8$ m/s, we can obtain $K$=0.2915=29.15%.

## VI. CALCULATION OF EFFECTIVE NONLINEAR SUSCEPTIBILITY

A classical expression for the intensity of the second harmonic generation (SHG) in a nonlinear medium is [3, 4]

$$I_{2\omega} = \frac{(2\omega)^2}{8c^3\varepsilon_0\varepsilon_\omega\sqrt{\varepsilon_{2\omega}}}|\chi_{eff}^{(2)}|^2 d^2 I_\omega^2 \cdot \text{sinc}^2(\frac{\Delta k d}{2}), \qquad (S17)$$

where $I_{2\omega}$ is the intensity of SHG, $I_\omega$ is the intensity of the incident wave at the angular frequency of $\omega$, $\varepsilon_0$ is the vacuum permittivity, $\Delta k = k_{2\omega} - 2k_\omega$, and $k_\omega$ and $k_{2\omega}$ are the wavenumbers of fundamental and second-order waves, $\varepsilon_\omega$ and $\varepsilon_{2\omega}$ are the relative permittivities of the nonlinear medium at frequency $\omega$ and $2\omega$ respectively, $\chi_{eff}^{(2)}$ is the effective nonlinear susceptibility, $d$ is the material thickness, and $c$ is the speed of light. Since its thickness is much thinner than the wavelength, the meta-molecule is equivalent to a very thin slab with a homogeneous material, and the refractive index at fundamental and second-order frequencies can be assumed as unity. Therefore, with $\Delta k$=0 and sinc($\Delta k d$/2)=1, the meta-molecule can be treated as a nonlinear material under phase-matching condition. In this case, the Eq. (S17) can be evolved as

$$I_{2\omega} = \frac{(2\omega)^2}{8c^3\varepsilon_0\varepsilon_\omega\sqrt{\varepsilon_{2\omega}}}\left|\chi_{eff}^{(2)}\right|^2 d^2 I_\omega^2. \qquad (S18)$$

This equivalent phase-matching condition of the meta-molecule can also be understood in the following way. The SHG intensity of the meta-molecule reaches maximum under the normal incidence with $x$-polarization as described in the manuscript. The commonly-used methods of achieving phase matching condition in nonlinear optics, such as oblique incidence and configuring the polarization of the incident wave, cannot improve the nonlinear response of the meta-molecule, since they may have negative impacts on the resonance or coupling of the meta-molecule. Considering the truth that natural nonlinear materials achieve the maximum intensity of SHG under the phase-matching condition, it is appropriate to assume the meta-molecule in an equivalent condition.

The intensity of the incident Gaussian wave can be expressed with the electric field as

$$I_\omega = 2\varepsilon_0 c\sqrt{\varepsilon_\omega}\left|\vec{E}(\omega)\right|^2. \qquad (S19)$$

Combing the Eqs. (S18) and (S19) with the permittivities of unity, we obtain the effective nonlinear susceptibility as

$$\left|\chi_{eff}^{(2)}\right| = \frac{c\sqrt{\eta}}{\omega d \left|\vec{E}(\omega)\right|}, \qquad (S20)$$

where $\eta = I_{2\omega}/I_\omega$ is the conversion efficiency of SHG. Since the nonlinear process originates from the coupling between the metal meta-atoms, the thickness of the material ($d$) should be that of the split-ring resonator (SRR) meta-atom. By substituting the speed of light and the parameters in the main text, including $\eta = 5.8\times10^{-9}$, $\left|\vec{E}(\omega)\right| = 10^7$ V/m, $\omega = 2\pi\times10^{10}$ rad/s, and $d=30$ μm, the effective nonlinear susceptibility can be worked out as $1.2\times10^3$ pm/V.

## VII. SIMULATION IN INFRARED REGIME

To verify the feasibility of the proposed theory of artificial nonlinearity in high frequency regime, we simulated a meta-molecule with the same structure as shown in Fig. (1) in the main text but working in the infrared regime. The geometric constants are scaled down to $l_1 = 1000$ nm, $w_1 = 180$ nm, $g=100$ nm, $l_2=600$ nm, $w_2=200$ nm, $d=20$ nm, and $P=1200$ nm. The substrate was 500 nm thick silicon dioxide with the permittivity of 4.82+0.026*i*, which is a more frequently-used substrate in the infrared regime than the FR-4. The SRR meta-atom is made of gold, which is more fabrication-friendly than copper in nanoscale, and the cut-wire meta-atom is made of bismuth. The thickness of the both meta-atoms are 100 nm. In the infrared regime, the metal materials are characterized by Drude model. For gold, the plasma frequency is $2\pi \times 2175$ THz with the collision frequency of $2\pi \times 6.5$ THz [5]. For bismuth, its collision frequency is $2\pi \times 0.75$ THz [6]. The approximation of $1-i\omega/\gamma \approx 1$ is not applicable, and a complex anisotropic conductivity tensor ($\tilde{\sigma}(\omega)$), based on Drude model as well, should be used

$$\tilde{\sigma}(\omega)=\tilde{\sigma}\begin{bmatrix} \dfrac{1}{1+(\tilde{\mu}_e B)^2} & -\dfrac{\tilde{\mu}_e B}{1+(\tilde{\mu}_e B)^2} & 0 \\ \dfrac{\tilde{\mu}_e B}{1+(\tilde{\mu}_e B)^2} & \dfrac{1}{1+(\tilde{\mu}_e B)^2} & 0 \\ 0 & 0 & 1 \end{bmatrix}, \qquad (S21)$$

where $\tilde{\sigma}=\dfrac{\sigma_0}{1-i\omega/\gamma}$ and $\tilde{\mu}_e=\dfrac{\mu_{e0}}{1-i\omega/\gamma}$ are the optical conductivity and mobility at frequency $\omega$, respectively. The dc conductivity ($\sigma_0$) and mobility ($\mu_{e0}$) are modeled with the $2.2\times10^5$ S/m and of 0.11 m$^2$/V•s respectively, same as the values used in the main text.

Under the normal illumination of an *x*-polarized plane wave from the top, the reflection, transmission and absorption spectra of the infrared meta-molecule were first simulated in the frequency domain and plotted in Fig. S3. The localized magnetic field reaches its maximum at 30 THz, which is slightly lower than the resonant frequency of 30.3 THz. The wavelength corresponding to 30 THz is 10 μm, belonging to the infrared regime, where the enhanced magnetic field is as maximum 15.9 times as the incident one.

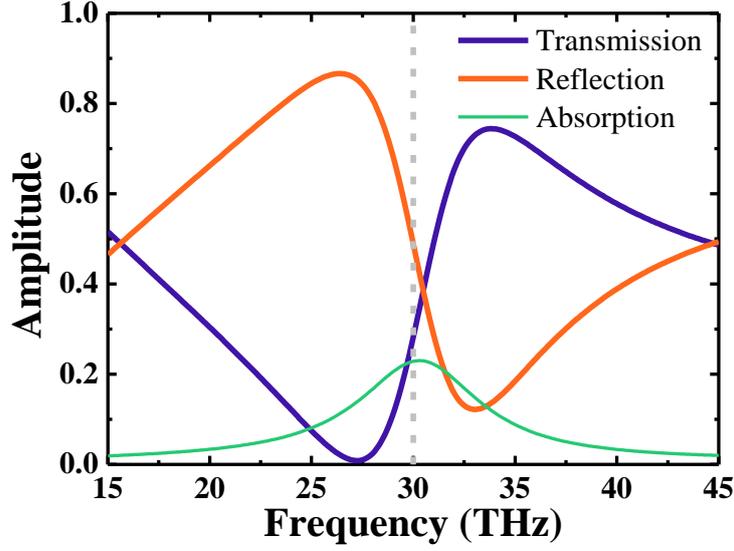

FIG. S3. Transmission, reflection and absorption spectra of the infrared meta-molecule with 30 THz marked with grey dashed line

The time-domain response of the meta-molecule was then simulated with a Gaussian pulsed plane wave at 30 THz casted from the top as well. The parameters for the Gaussian wave are $|\vec{E}(\omega)|=10^7$ V/m, $\omega=2\pi\times30\times10^{12}$ rad/s, $t_0=800$ fs, $\Delta t=300$ fs, and the total time of 2 ps was simulated with the step of 1 fs. As revealed in Fig. S4(a), a 60 THz wave can be extracted from the $y$-polarized transmission spectrum by high-pass filtering with the peak electric field of 47.6 V/m. The doubled frequency demonstrates the successful generation of the second-harmonic wave. The frequency spectrum in Fig. S4(b), transformed from the time spectrum by Fourier transformation, exhibits the SHG more obviously.

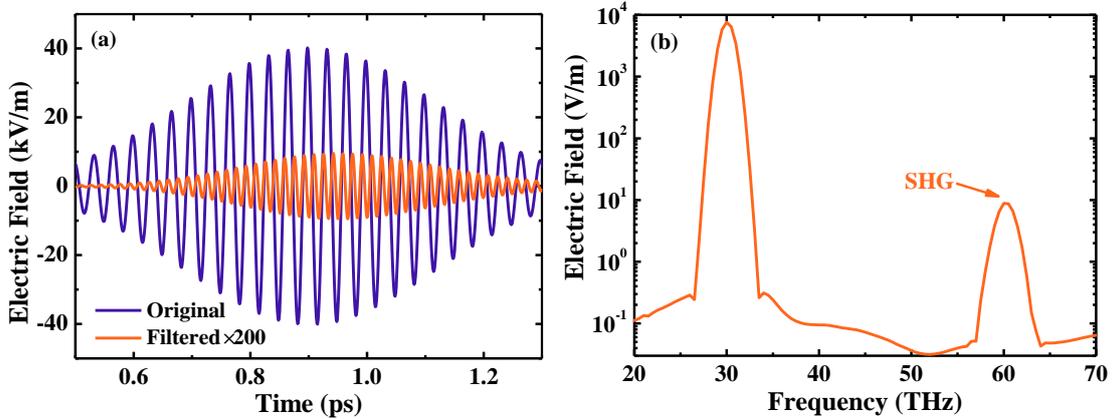

FIG. S4. Original and high-pass filtered time-domain transmission spectra of the infrared meta-molecule in $y$ polarization (a) and its frequency-domain spectrum in logarithmic coordinate (b).

The reflection spectrum also presents distinct SHG with the peak electric field of 22.5 V/m (not shown). Therefore, the effective nonlinear susceptibility of 8.4 pm/V can be calculated with the same method described in the section V. Despite the smaller susceptibility in the infrared regime than that in the microwave regime, the simulated results fully demonstrate the proposed theory of artificial nonlinearity is feasible in the infrared regime. We believe the relatively weak nonlinear response results from the dispersive characteristics of the materials at high frequency, which leads to the decrease of the localized magnetic field induced by the SRR meta-atom and the drift velocity of the free electrons in the cut-wire meta-atom.